\documentclass[12pt]{article} 
\usepackage{amsmath,amsthm,amssymb,bm,mathrsfs}
\usepackage{amscd}
\usepackage{graphicx}
\usepackage{typearea}
\typearea{12}

\newcommand{\Coef}[1]{\ensuremath{C_{\mathrm{#1}}}} 
\newcommand{\opt}[1]{\ensuremath{#1_{\mathrm{opt}}}} 
\newcommand{\sub}[2]{\ensuremath{#1_{\mathrm{#2}}}} 

\title{Projection Angles of Projectiles in Sports
\thanks{This manuscript is an English translation of a conference paper 
originally written in Japanese, and 
the original paper was presented 
at Sports Engineering and Human Dynamics 2025 (SHD2025), 
organized by the Japan Society of Mechanical Engineers \cite{Tsuboi_2025}.}\\
\large Qualitative Assessment of the Effects of Aerodynamic Forces or Run-Up
}

\author{Kazuhiro TSUBOI
\thanks{Professor Emeritus, Ibaraki University.
Email: kazuhiro.tsuboi.508@vc.ibaraki.ac.jp}\\
Graduate School of Sci. and Eng., Ibaraki University, Japan}

\date{ver. 1 in September 2026}

\begin{document}
\maketitle

\begin{abstract}
The determination of optimum projection angles is a fundamental problem 
in the mechanics of sports projectiles. In this study, we examine two major factors 
that influence the optimum angle: aerodynamic forces and the effect of run-up. 
With respect to aerodynamics, we consider not only the drag 
but also the lift generated by spin during flight. 
By linearizing the equations of motion that include these forces, 
we derive perturbative solutions with respect to drag and lift coefficients 
and clarify their qualitative effects. The results show that both drag 
and lift reduce the optimum projection angle, 
with the latter exerting a stronger influence. 
To investigate the effect of run-up, we propose an extended projection model 
in which the initial speed depends on the projection angle. 
Analysis of this model reveals that a stronger run-up increases 
the relative projection angle but decreases the launch angle observed from the ground. 
These findings provide a mechanical explanation for the throwing angle in shot put 
and the takeoff angle in long jump. 
The present study establishes a simple theoretical framework for clarifying 
the respective roles of aerodynamic and run-up effects 
in determining the optimum projection angles in sports. 
\end{abstract}


\section{Introduction}

Projectile motion is encountered in a wide variety of sports 
involving balls, throwing events, and jumping events \cite{Clanet}. 
In ball games such as baseball, soccer, and golf, 
the flight of the ball largely affects the outcome of play. 
In throwing events, including the shot put, javelin throw, and hammer throw, 
athletes seek to maximize the flight distance of projectiles. 
Even in jumping events such as the long jump, triple jump, and high jump, 
the athlete's body itself behaves as a projectile during the aerial phase.

In these sports, the projection angle is one of the most important factors 
governing projectile motion. 
Knowledge of the optimum projection angle is therefore essential 
not only for maximizing the flight distance 
but also for controlling the flight trajectory according to tactical or technical requirements.

The problem of determining the optimum projection angle has been studied 
since the early development of ballistics in the Middle Ages \cite{Stewart}. 
It is well known that, in the absence of external forces other than gravity, 
the optimum projection angle is $45^\circ$. 
In actual sports, however, considerably smaller projection angles are commonly observed. 
Typical examples include launch angles of approximately $25^\circ$--$30^\circ$ 
for baseball batting, around $10^\circ$ for golf drives \cite{Tait}, 
approximately $30^\circ$--$40^\circ$ for the shot put \cite{Linthorne_sp}, 
and takeoff angles of about $20^\circ$--$30^\circ$ 
in the long jump \cite{Linthorne_lj}.

Since projectile motion is governed by the laws of mechanics, 
these observed projection angles should be explainable from a mechanical viewpoint. 
The present study focuses on two principal factors 
that determine the optimum projection angle: 
aerodynamic forces acting in flight and the initial conditions 
associated with run-up before projection. 
For aerodynamic effects, both drag and lift 
generated by ball spin are taken into account. 
Their qualitative effects on the optimum projection angle 
are investigated by means of perturbation solutions derived 
from linearized equations of motion. 
To examine the effect of run-up, an extended projection model 
is introduced in which the projection speed depends on the projection angle. 
The optimum projection angle predicted by this model is then analyzed and discussed.

\section{Projectile Motion and Optimum Projection Angle}

\subsection{Equations of Motion for a Projectile Subject to Aerodynamic Forces}

We consider two-dimensional projectile motion in the $x$-$y$ plane. 
The origin of the coordinate system is taken at the center of mass of the projectile 
at the instant of projection, with the $y$-axis directed vertically upward 
and the $x$-axis lying in the horizontal direction within the plane defined 
by the initial velocity vector and the vertical axis.

Suppose that the projectile is subjected to drag and lift forces proportional to 
the square of its speed. 
The equations of motion of the center of mass 
and the corresponding initial conditions are given by
\begin{equation}
m\frac{du}{dt} = - kqu - lqv,\quad m\frac{dv}{dt} = - kqv + lqu - mg,
\label{eq:1}
\end{equation}
\begin{equation}
u(0)=q_i \cos\theta \equiv u_i,\quad v(0)=q_i \sin\theta \equiv v_i,\quad
x(0) = y(0) = 0,
\label{eq:2}
\end{equation}
where $(x,y)$ and $(u,v)$ denote the position and velocity vectors of the projectile, 
respectively, and $q^2=u^2+v^2$ is the squared magnitude of the velocity. 
Furthermore, $m$ denotes the projectile mass, $g$ the gravitational acceleration, 
while $q_i$ and $\theta$ represent the initial speed and the projection angle, respectively.

The proportional constants $k$ and $l$ appearing in the terms of drag and lift
in Eq.~\eqref{eq:1}, are defined by
\begin{equation}
k = \frac{1}{2}\rho A \Coef{D},\quad l = \frac{1}{2}\rho A \Coef{L},
\label{eq:3}
\end{equation}
where $\rho$ is the air density, $A$ is the cross-sectional area of the projectile, 
and \Coef{D} and \Coef{L} are the drag and lift coefficients, respectively.

Using the initial speed $q_i$ and the gravitational acceleration $g$ 
as characteristic quantities, all quantities in the governing equations 
can be nondimensionalized as follows:
\[
u = u^* q_i,\quad v = v^* q_i,\quad q = q^* q_i,\quad
t = t^* \frac{q_i}{g},\quad 
x = x^* \frac{q_{i}^2}{g},\quad y = y^* \frac{q_{i}^2}{g},
\]
where the quantities marked by $*$ denote dimensionless variables.

The equations of motion \eqref{eq:1} and the initial conditions \eqref{eq:2} 
are then rewritten as
\begin{equation}
\frac{du^*}{dt^*} = - \sub{\varepsilon}{D} q^* u^* - \sub{\varepsilon}{L} q^* v^*,\quad 
\frac{dv^*}{dt^*} = - \sub{\varepsilon}{D} q^* v^* + \sub{\varepsilon}{L} q^* u^* - 1,
\label{eq:4}
\end{equation}
\begin{equation}
u^*(0)=\cos\theta \equiv u^{*}_i,\quad v^*(0)=\sin\theta \equiv v^{*}_i,\quad
x^*(0) = y^*(0) = 0,
\label{eq:5}
\end{equation}
where the dimensionless parameters \sub{\varepsilon}{D} 
and \sub{\varepsilon}{L} are defined by
\begin{equation}
\varepsilon_D = \frac{kq_{i}^2}{mg},\quad 
\sub{\varepsilon}{L} = \frac{lq_{i}^2}{mg}.
\label{eq:6}
\end{equation}

The parameters \sub{\varepsilon}{D} and \sub{\varepsilon}{L} are referred to 
as the drag-to-weight ratio and the lift-to-weight ratio, respectively. 
For typical sports projectiles, the drag-to-weight ratio is approximately 
in the range of $0.01 \le \sub{\varepsilon}{D} \le 1$,
as reported in \cite{de Mestre}. 

In the remainder of this paper, 
all quantities are assumed to be dimensionless unless otherwise stated, 
and the superscript $*$ indicating nondimensional variables 
is omitted for simplicity.

\subsection{Derivation Procedure of the Optimum Projection Angle}

The general procedure for determining the optimum projection angle of a point mass projectile 
is summarized as follows \cite{Tsuboi_2015}.

\medskip

\noindent
\textbf{Step 1.}
Obtain the projectile position by solving the equations of motion \eqref{eq:4} 
subject to the initial conditions \eqref{eq:5}: 
\[
x=x(t,\sub{\varepsilon}{D},\sub{\varepsilon}{L},u_i,v_i), \qquad
y=y(t,\sub{\varepsilon}{D},\sub{\varepsilon}{L},u_i,v_i).
\]
\noindent
\textbf{Step 2.}
Eliminating the time variable $t$ from the above solutions,  
derive the trajectory solution 
$y=y(x,\sub{\varepsilon}{D},\sub{\varepsilon}{L},u_i,v_i).$

\medskip
\noindent
\textbf{Step 3.}
Determine the flight range 
$L=L(\zeta,\sub{\varepsilon}{D},\sub{\varepsilon}{L},u_i,v_i)$ 
from the solution of the equation 
$y(L,\sub{\varepsilon}{D},\sub{\varepsilon}{L},u_i,v_i)=\zeta$.
Here $\zeta$ denotes the vertical displacement 
between the projection and landing positions, 
and the conservation of mechanical energy implies 
the condition $\zeta \le 1/2$.

In particular, when the trajectory solution cannot be expressed explicitly, 
the flight time $T=T(\zeta,\sub{\varepsilon}{D},\sub{\varepsilon}{L},u_i,v_i)$ is determined 
from the solution of the equation in the vertical position 
$y(T,\sub{\varepsilon}{D},\sub{\varepsilon}{L},u_i,v_i)=\zeta$,
and substituting the flight time $T$ into the horizontal position, 
we can obtain the flight range 
$L=x(T,\sub{\varepsilon}{D},\sub{\varepsilon}{L},u_i,v_i)$.
\medskip

\noindent
\textbf{Step 4.}
Through the initial velocity vector $(u_i, v_i)$, 
the flight range becomes a function of the projection angle $\theta$, 
that is 
$L=L(\zeta,\sub{\varepsilon}{D},\sub{\varepsilon}{L},\theta)$. 
Thus the optimum projection angle \opt{\theta} is determined 
from the following stationary condition
\begin{equation}
\frac{dL}{d\theta} = 0.
\label{eq:7}
\end{equation}

Equation \eqref{eq:4} is nonlinear, and the projectile position 
cannot be obtained explicitly as a function of time.
We therefore introduce the parameters
\begin{equation}
\alpha \equiv \sub{\varepsilon}{D} q \approx \sub{\varepsilon}{D},\quad
\beta  \equiv \sub{\varepsilon}{L} q \approx \sub{\varepsilon}{L},
\label{eq:8}
\end{equation}
and assume them to be constant, 
thereby obtaining a linearized form of the governing equations,
\begin{equation}
\frac{du}{dt} = -\alpha u - \beta v,\quad
\frac{dv}{dt} = -\alpha v + \beta u - 1.
\label{eq:9}
\end{equation}

When the projectile speed satisfies $q=1$, the parameters $\alpha$ and $\beta$ 
coincide with \sub{\varepsilon}{D} and \sub{\varepsilon}{L}, respectively.
Consequently, the approximation introduced in Eq.~\eqref{eq:8} corresponds 
to assuming that the projectile speed appearing 
in the aerodynamic forces remains equal to its initial value throughout the flight.

Although the linearized equations of motion have explicit solutions 
for the projectile position, the trajectory solution cannot be obtained analytically 
in terms of the aerodynamic parameters and the initial velocity.
Accordingly, the subsequent analysis is carried out using perturbation expansions 
with respect to the aerodynamic parameters\cite{Tsuboi_2012}.

Let $L_0$ denote the zeroth-order approximation of the flight range $L$ 
corresponding to projectile motion in vacuum. 
The zeroth-order approximation of the stationary condition is then 
obtained from Eq.~\eqref{eq:7} as follows, 
\[
\frac{dL_0}{d\theta} = \frac{d}{d\theta} 
\left[ u_i \left( v_i + \sqrt{v_{i}^2 - 2\zeta} \right) \right],
\]
which yields the zeroth-order solution of the optimum projection angle.

After straightforward manipulation, 
the stationary condition to be solved can be written as
\begin{equation}
\left( u_i v_{i}' \right)^2 - \left( u_{i}' v_{i} \right)^2 
+ 2\zeta \left(u_{i}'\right)^2 = 0.
\label{eq:10}
\end{equation}
where the prime $'$ denotes differentiation with respect to $\theta$.

\section{Effect of Aerodynamic Forces}

\subsection{Perturbation Solution}

As mentioned in the previous section, the trajectory solution required in Step 2 
cannot be obtained explicitly from the solution of the linearized equations of motion. 
Consequently, the flight range is determined from the flight time according to the procedure described in Step 3.

Since the flight time cannot be expressed analytically, 
perturbation expansions with respect to the aerodynamic parameters 
$\alpha$ and $\beta$ are employed here. 
Assuming, for simplicity, that $\zeta = 0$, 
the flight time $T$ and the flight range $L$ are obtained 
up to the second order as follows:
\begin{align}
\begin{split}\label{eq:11}
T &= 2 v_i + \alpha \left( -\frac{2}{3} v_{i}^2 + 2 u_i v_i r \right)\\
  & \quad + \alpha^2 \left[\frac{4}{9} - \frac{8}{3} u_i v_{i}^2 r  
+  2 \left( u_{i}^2 v_i - \frac{1}{3} v_{i}^3 \right) r^2 \right],
\end{split}\\
\begin{split}\label{eq:12}
L &= 2 u_i v_i + \alpha \left( -\frac{8}{3} u_i v_{i}^2 
+   \frac{2}{3} v_i \left( 3 u_{i}^2 - v_{i}^2 \right) r \right)\\
  & \quad + \alpha^2 \left(\frac{28}{9} u_i v_{i}^3 - \frac{4}{3} v_{i}^2 \left( u_{i}^2 - v_{i}^2\right) r  
+ 2 u_i v_i \left( u_{i}^2 - v_{i}^2 \right) r^2 \right),
\end{split}
\end{align}
where $r=\beta/\alpha$ denotes the ratio of lift to drag.

Applying the stationary condition in Step 4 to the perturbation solution 
for the range yields the following perturbation solution 
for the optimum projection angle,
\begin{align}
\label{eq:13}
\opt{\theta} = 
\frac{\pi}{4} - \alpha\left( \frac{\sqrt{2}}{6}+\frac{\sqrt{2}}{4}r \right) 
+ \alpha^2\left(\frac{1}{9} + \frac{1}{6}r - \frac{1}{8}r^2 \right).
\end{align}

The perturbation solution predicts the following qualitative characteristics 
of the optimum projection angle \cite{Tsuboi_2012}.
\begin{enumerate}
\item
The signs of the first-order coefficients indicate that both drag and lift 
reduce the optimum projection angle.
\item
The first-order coefficients further show that lift has a stronger influence 
than drag. In the linearized model, 
when the aerodynamic forces are sufficiently small, 
the reduction due to lift is approximately 1.5 ($= (\sqrt{2}/4)/(\sqrt{2}/6)$)
times larger than that due to drag.
\item
The signs of the second-order coefficients indicate 
that the optimum projection angle is a concave function of the drag parameter, 
whereas it is a convex function of the lift parameter.
\end{enumerate}

\subsection{Effect of Drag Alone}

We first examine the effect of drag alone by considering the case 
in which no lift acts on the projectile. 
The variation of the optimum projection angle \opt{\theta} with the drag parameter 
is shown in Figure~1. 
Throughout this section, the linearization introduced in Eq.~\eqref{eq:8} 
is adopted so that $\sub{\varepsilon}{D}=\alpha$ and $\sub{\varepsilon}{L}=\beta$.

In Figure~1, the solid curve represents the second-order perturbation solution 
obtained from Eq.~\eqref{eq:13} with $r=0$, 
whereas the diamond symbols $\diamond$ denote numerical solutions 
of the linearized equations of motion \eqref{eq:9}. 
The numerical solutions were obtained using the fourth-order Runge--Kutta method 
with a time step of $\Delta t=10^{-3}$. 
The projection angle was varied from $\pi/4$ in increments of 
$\pi/100$ (=$1.8^\circ$), and the optimum angle was determined 
as the angle giving the maximum horizontal range.

The results show that the optimum projection angle decreases 
to approximately $35^\circ$ when $\alpha\,(=\sub{\varepsilon}{D})\approx1.0$, 
indicating that stronger drag leads to a smaller optimum projection angle. 
Comparison with the numerical solutions further demonstrates 
that the perturbation solution given by Eq.~\eqref{eq:13} remains accurate 
over the range $\sub{\varepsilon}{D}(=\alpha) \lesssim 0.5$. 
Moreover, both the perturbation and numerical solutions exhibit 
a concave dependence on the drag parameter, showing qualitative agreement.
\begin{figure}[htbp]
\centering
\includegraphics[width=.6\linewidth]{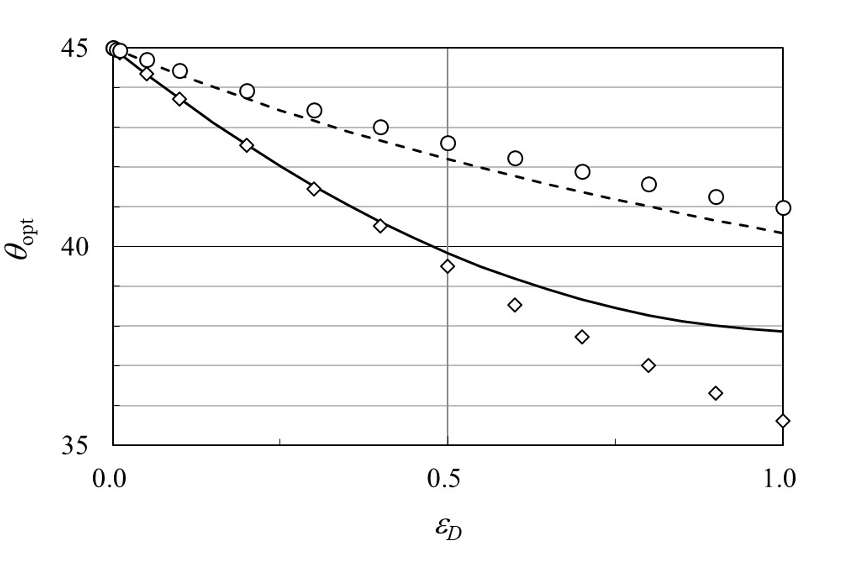}
\caption{Effect of the drag coefficient on the optimum projection angle. 
The solid line shows the perturbation solution \eqref{eq:13} 
with $r = 0$, and the diamond symbols $\diamond$ denote numerical solutions of Eq.~\eqref{eq:9}. 
The broken line represents the approximate solution \eqref{eq:14}, 
and the open circles $\circ$ correspond to numerical solutions of Eq.~\eqref{eq:4}.}
\label{fig:1}
\end{figure}

To examine the effect of the quadratic drag law, 
the corresponding numerical solutions of Eq.~\eqref{eq:4} are also plotted 
in Figure~1 as open circles $\circ$. 
For comparison, the approximate solution proposed 
by \cite{Cohen},
\begin{align}
\label{eq:14}
\tan\theta = \sqrt{\frac{\sub{\varepsilon}{D}}
{\left( 1 + \sub{\varepsilon}{D} \right)\log\left( 1 + \sub{\varepsilon}{D} \right)}},
\end{align}
is shown by the broken curve.
 
The quadratic drag law also predicts 
that the optimum projection angle decreases monotonically 
with increasing drag coefficient 
while exhibiting a concave dependence on the drag parameter. 
Thus, the qualitative behavior agrees well with that obtained 
from the linearized model.

Quantitatively, however, the reduction predicted by the linearized model 
is considerably larger. 
Under the quadratic drag law, the optimum angle remains 
approximately $41^\circ$ at $\sub{\varepsilon}{D}=1$, 
whereas the linearized model predicts about $35^\circ$. 
In the range of $\sub{\varepsilon}{D}\ll1$, the initial slope obtained 
from the linearized model is approximately 2.6 ($=(\sqrt{2}/6)/(0.08994)$) times 
larger than that of the quadratic drag law \cite{Tsuboi_2012}. 
Consequently, the reduction of the optimum angle at $\sub{\varepsilon}{D}=1$ 
is approximately $4^\circ$ for the quadratic drag law 
but about $9^\circ$ for the linearized model.

This discrepancy originates from the approximation introduced in Eq.~\eqref{eq:8}. 
In the quadratic drag law, the projectile speed gradually decreases 
during the flight because of aerodynamic drag, 
causing the parameter $\alpha$ to decrease accordingly. 
In the linearized model, however, $\alpha$ is assumed to remain constant 
throughout the motion. 
As a result, the drag force is overestimated, 
leading to a larger reduction in the optimum projection angle.

\subsection{Effect of Lift Alone}

We next consider the case in which drag is absent ($\sub{\varepsilon}{D}=\alpha=0$). 
Although this situation is not physically realizable, 
it admits an exact analytical solution 
and therefore provides valuable insight into the role of lift 
in determining the optimum projection angle.

Under this assumption, 
the flight time and the flight range appearing in Step~3 are obtained as
\begin{align}
\label{eq:15}
\begin{split}
\tan\beta T &= - \frac{2\beta\sin\theta\left( \beta\cos\theta - 1 \right)}
{\left( \beta\cos\theta - 1 \right)^2 - \beta^2 \sin^2\theta}, \\
L &= -\frac{1}{\beta^2}
\left[ 2\beta\sin\theta + \tan^{-1}
\left( \frac{2\beta\sin\theta \left(\beta\cos\theta -1\right)}
            {\left(\beta\cos\theta -1\right)^2 - \beta^2 \sin^2 \theta} 
\right)
\right].
\end{split}
\end{align}

The stationary condition in Step~4 then yields the following exact solution 
for the optimum projection angle,
\begin{align}
\label{eq:16}
\cos\opt{\theta} &= \frac{1}{4}\left( \beta + \sqrt{\beta^2 + 8} \right),
\end{align}

Figure~2 shows the variation of the optimum projection angle 
with the lift parameter $\sub{\varepsilon}{L}\,(=\beta)$. 
The solid curve represents the exact solution given by Eq.~(16), 
while the broken curve denotes the second-order perturbation solution 
obtained from Eq.~(13) with $\alpha=0$.

\begin{figure}[htbp]
\centering
\includegraphics[width=.6\linewidth]{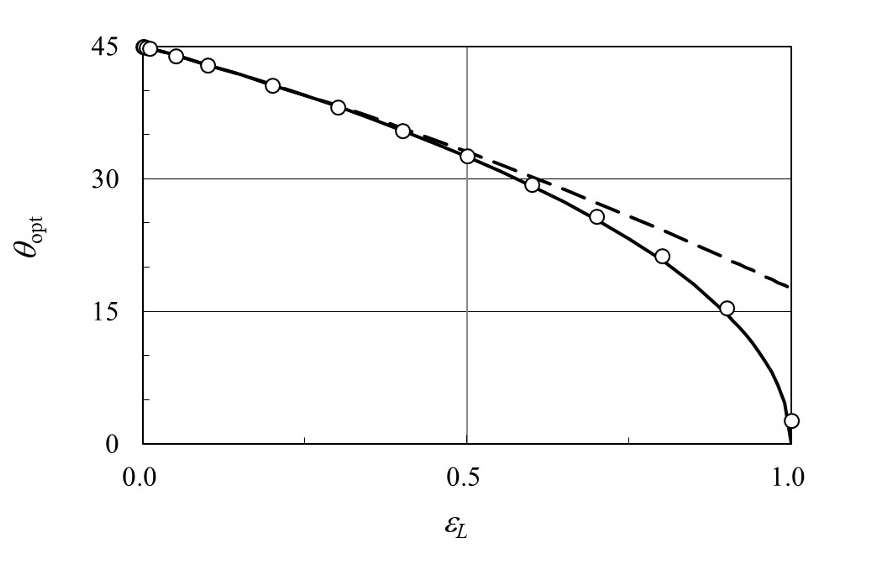}
\caption{Effect of the lift coefficient on the optimum projection angle. 
The solid line indicates the exact solution \eqref{eq:16}, 
and the broken line denotes the second-order perturbation solution \eqref{eq:13} 
with $\alpha = 0$. 
The open circles $\circ$ correspond to numerical solutions 
based on the quadratic law of lift force.}
\label{fig:2}
\end{figure}

Both solutions indicate that the optimum projection angle 
is a convex function of the lift parameter. 
Consequently, the optimum angle decreases rapidly as the lift coefficient increases. 
According to the exact solution, 
the optimum projection angle reaches $0^\circ$ at $\sub{\varepsilon}{L}\,(=\beta)=1.0$. 
Comparison with the perturbation solution shows 
that the second-order approximation remains valid 
over the range of $\sub{\varepsilon}{L}(=\beta)\lesssim0.5$.

The open circles $\circ$ in Figure~2 represent numerical solutions of Eq.~(4) 
obtained under the quadratic lift law with $\sub{\varepsilon}{D}=0$.

An interesting observation is that the numerical solutions 
based on the quadratic lift law agree remarkably well 
with the exact solution of the linearized equations, 
despite the different velocity dependence of the lift force. 
At present, however, the reason for this agreement remains unknown.

\subsection{Combined Effects of Drag and Lift}

Figure~3 summarizes the dependence of the optimum projection angle 
on both the drag-to-weight ratio $\sub{\varepsilon}{D}\,(=\alpha)$ 
and the lift-to-drag ratio $r$, 
based on the perturbation solution given by Eq.~(13). 
The curve corresponding to $r=0$ and the open circles 
are identical to those presented in Figure~1. 
In addition, the diamond symbols represent numerical solutions 
of the linearized equations of motion (9) for the case $r=1.0$ ($\alpha=\beta$). 
As in the previous sections, the second-order perturbation solution 
is found to provide an accurate approximation 
over the range of $\sub{\varepsilon}{D}(=\alpha)\lesssim0.5$.

As shown in Section~3.2, drag reduces the optimum projection angle 
along a concave curve, so that the rate of decrease becomes 
gradually smaller as the drag parameter increases. 
In contrast, we demonstrated in Section~3.3 that lift decreases 
the optimum projection angle along a convex curve, 
leading to an increasingly rapid reduction as the lift parameter becomes larger.

Figure~3 clearly illustrates the interaction between these two aerodynamic effects. 
For small values of the lift-to-drag ratio, the effect of drag predominates, 
and the decrease in the optimum projection angle gradually weakens. 
As the lift-to-drag ratio increases, however, 
the contribution of lift becomes more significant, 
causing the curves to approach a nearly linear behavior 
and resulting in a much larger reduction of the optimum projection angle.

Although the results presented in Figure~3 are 
based on the linearized equations of motion 
and therefore underestimate the optimum projection angle quantitatively, 
the qualitative dependence on the aerodynamic parameters 
is expected to remain valid for the quadratic aerodynamic law as well.

In particular, the results in Section~3.3 imply 
that, in the range of $\sub{\varepsilon}{D}\ll1$, the initial rate of decrease 
of the optimum projection angle due to lift 
under the quadratic law is $-\sqrt{2}/4$ from Eq.~\eqref{eq:13}. 
Consequently, the ratio of the lift-induced reduction to 
that caused by drag is approximately 3.9 ($=(\sqrt{2}/4/(0.08994)$).

\begin{figure}[htbp]
\centering
\includegraphics[width=.6\linewidth]{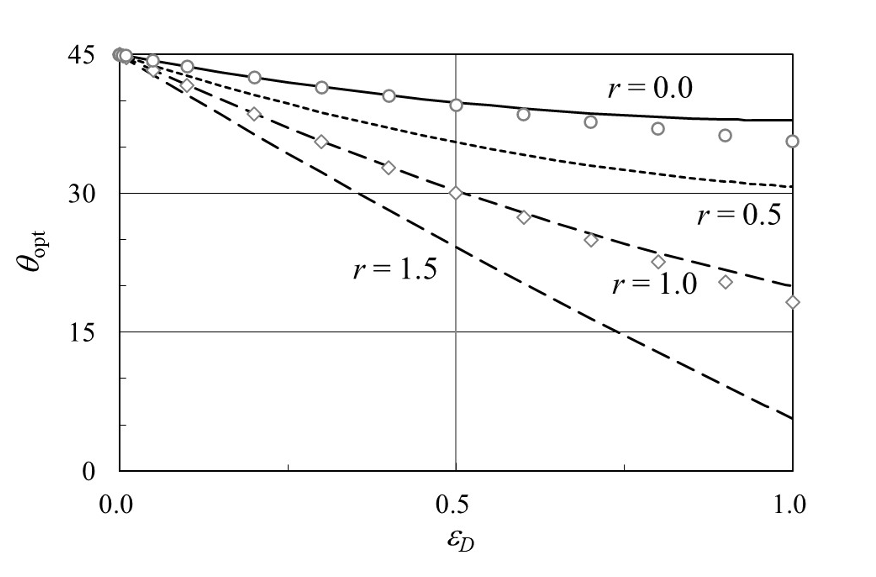}
\caption{Effect of aerodynamic coefficients on the optimum projection angle. 
The perturbation solution \eqref{eq:13} is shown 
for each value of the lift-to-drag ratio $r$. 
The open circles $\circ$ and diamond $\diamond$ symbols represent 
numerical solutions of the linearized equation \eqref{eq:9}.}
\label{fig:3}
\end{figure}

\section{Effect of Run-Up}

\subsection{Extended Projection Model}

Figure~4 illustrates a simple projection model 
in which a point mass is projected with speed $w$ at an angle $\psi$ 
from a platform moving horizontally at a constant speed $V\,( \ge 0 )$. 
This model provides the simplest mechanical representation of projection motion 
accompanied by a run-up, as commonly observed in sports \cite{Tsuboi_2010}.

\begin{figure}[htbp]
\centering
\includegraphics[width=.75\linewidth]{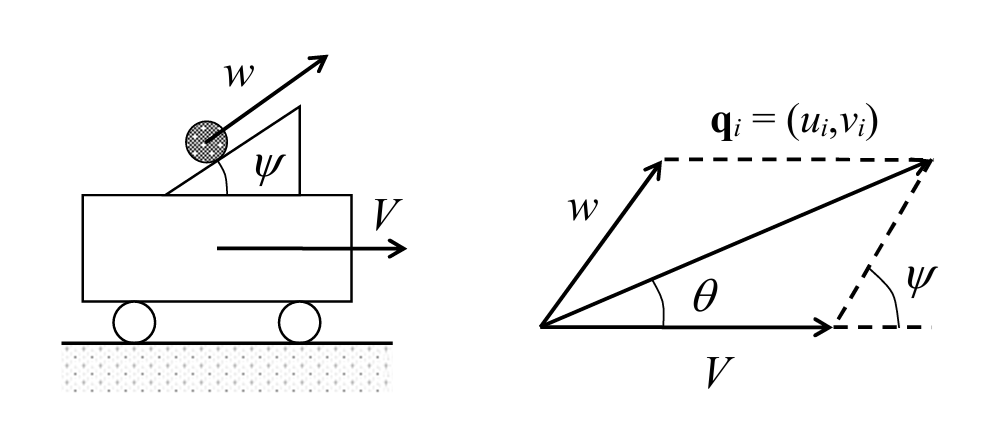}
\caption{Schematic representation of the extended projection model. 
In this model, a point mass is projected with speed $w$ at an angle $\psi$ 
from a platform moving at constant speed $V$ along the horizontal plane (left). 
The initial velocity is given by the resultant of $w$ and $V$ (right).}
\label{fig:4}
\end{figure}

The horizontal motion of the platform represents 
the translational motion of the athlete's upper body or 
trunk generated during the run-up, 
whereas the projection velocity relative to the platform 
is produced by the action of other body segments, such as the arms or legs.

The initial velocity of the projectile with respect to the ground 
is therefore given by the vector sum of the platform velocity $V$ 
and the relative projection velocity $w$, yielding
\begin{align}
\label{eq:17}
u_i = q_i(\theta)\cos\theta = V + w\cos\psi, \qquad
v_i = q_i(\theta)\sin\theta = w\sin\psi,
\end{align}
where $q_i^2=u_i^2+v_i^2$ denotes the squared magnitude of the initial velocity.

In this section, $\psi$ is referred to as the projection angle 
in order to distinguish it from the initial angle $\theta$. 
Since the objective is to maximize the horizontal range, only
$0\le\psi\le90^\circ$ and $\ 0\le\theta\le90^\circ$ 
are considered.

In ordinary projection motion, the initial velocity vector 
is completely specified by its magnitude $q_i$ and direction $\theta$. 
In contrast, the present projection model introduces three independent variables, 
namely the projection speed $w$, the platform speed $V$, 
and the projection angle $\psi$.

The conventional projection model is recovered in the limiting case
$V\rightarrow0$.
For this reason, the present model will hereafter be referred to 
as the extended projection model.

From the velocity triangle shown in Figure~4, 
the magnitude of the initial velocity can be expressed 
as a function of the initial angle,
\begin{align}
\label{eq:18}
q_i(\theta) = V\cos\theta + \sqrt{w^2 - V^2 \sin^2\theta}.
\end{align}

This expression immediately shows that the initial speed 
is a monotonically decreasing function of the initial angle. 
This dependence of the initial speed on the initial angle constitutes 
the essential mechanical feature of projection motion with run-up \cite{Tsuboi_2015}.

\subsection{Optimum Projection Angle}

In this section, our attention is focused on 
the dependence of the initial speed on the projection angle, 
and hence the aerodynamic forces acting in aerial phase are neglected.

Substituting the initial velocity given by Eq.~\eqref{eq:17} 
into the zeroth-order stationary condition \eqref{eq:10}, 
we obtain the following cubic equation for the projection angle $\psi$,
\begin{align}
\label{eq:19}
2 \gamma \cos^3\psi + \left( \gamma^2 + 2 - 2\zeta \right)\cos^2\psi + 2\zeta-1 = 0,
\end{align}
where $\gamma=V/w$ denotes the speed ratio, 
which characterizes the degree of run-up.

A simple analysis shows that this cubic equation has three real roots. 
Application of the solution formula in cubic equation yields the following solution 
corresponding to the optimum projection angle $\psi$ \cite{Tsuboi_2010},
\begin{align}
\label{eq:20}
\cos\psi = \frac{\gamma^2 + 2 -2\zeta}{6\gamma} \left( 2\cos\frac{\varphi}{3} -1 \right), 
\qquad
\cos\varphi = 54 \frac{\gamma^2 \left( 1 - 2\zeta \right)}{\left( \gamma^2 + 2 -2\zeta \right)^3} - 1.
\end{align}

The corresponding optimum initial angle $\theta$ is then obtained 
from the geometrical relationship between the projection angle and the initial angle,
\begin{align}
\label{eq:21}
\tan\theta = \frac{\sin\psi}{\gamma+\cos\psi}.
\end{align}

Figure~5 shows the optimum projection angle \opt{\psi}
and the optimum initial angle \opt{\theta}
as functions of the speed ratio $\gamma$.
The broken curves represent \opt{\psi}, 
whereas the solid curves denote \opt{\theta}. 
Different colors correspond to different vertical displacement  
between the projection and landing points: blue, black, red, and green 
indicate $\zeta=0.25$, $0.0$, $-0.5$, and $-1.0$, respectively.

As is evident from Figure~5, the optimum projection angle \opt{\psi} increases rapidly 
over the range $0.1\le\gamma\le10$, almost independently of the vertical displacement. 
In contrast, the optimum initial angle \opt{\theta} decreases rapidly over the same range.

Although the effect of the vertical displacement is significant 
when $\gamma<0.1$, it gradually diminishes with increasing $\gamma$. 
For sufficiently large values of $\gamma$, 
the optimum projection angle and the optimum initial angle 
asymptotically approach $90^\circ$ and $0^\circ$, respectively.
These limiting behaviors are consistent with the fact 
that the present model approaches ordinary projection model 
in the limit of small $\gamma$.

The results shown in Figure~5 indicate 
that increasing the horizontal velocity associated 
with the run-up increases the optimum projection angle 
while simultaneously decreasing the optimum initial angle. 
This behavior represents a fundamental characteristic of projection motion 
accompanied by run-up.

The model also provides a simple mechanical explanation 
for the release angle in the shot put 
and the takeoff angle in the long jump. 
Numerical estimation based on experimental measurements indicate 
that the speed ratio is approximately $0.4\le\gamma\le0.8$ for the shot put 
and $1.4\le\gamma\le1.8$ for the long jump, 
while the vertical displacement is 
approximately $\zeta=-0.3$ in both events \cite{Tsuboi_jsiam_2015}.

Figure 5 therefore predicts optimum initial angles of 
approximately $30^\circ$--$35^\circ$ for the shot put 
and $20^\circ$--$25^\circ$ for the long jump,
in good agreement with experimental observations.
\begin{figure}[hbtp]
\centering
\includegraphics[width=.6\linewidth]{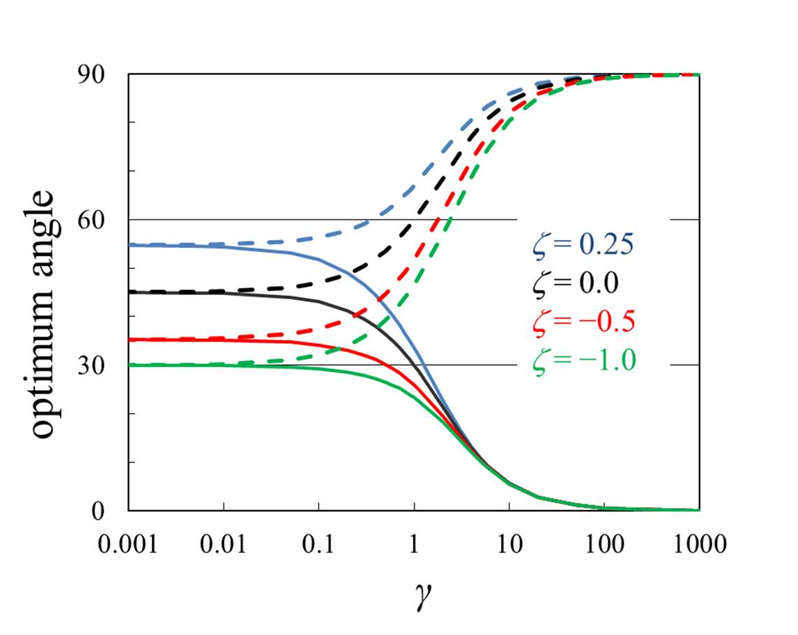}
\caption{Optimum projection and initial angles in the extended projection model. 
The solid and broken lines represent the results for the initial angle \opt{\theta}
and the projection angle \opt{\psi}. 
Colors indicate different values of $\zeta$: 
blue, black, red and green  correspond to 
$\zeta = 0.25, 0.0, -0.5$ and $-1.0$, respectively. 
The optimum initial angle \sub{\theta}{opt} exhibits a drastic decrease 
in the range of $0.1 \le \gamma \le 10$.}
\label{fig:5}
\end{figure}

\section{Conclusions}

The optimum projection angle of sports projectiles has been investigated 
from two complementary viewpoints: 
aerodynamic forces acting in flight and the effect of run-up before projection.

For the aerodynamic effects, perturbation solutions were derived 
from linearized equations of motion including drag and lift. 
The analysis demonstrated that both aerodynamic forces reduce 
the optimum projection angle, 
while the effect of lift is greater than that of drag. 
Although the linearized model overestimates the quantitative effect of drag, 
it successfully reproduces the qualitative dependence of the optimum projection angle 
on the aerodynamic coefficients.

To investigate the effect of run-up, an extended projection model was used 
in which the projection speed depends on the projection angle. 
The analysis showed that increasing the run-up raises the projection angle 
relative to the athlete 
while simultaneously lowering the initial launch angle measured from the ground. 
The model provides a simple mechanical explanation 
for the optimum release angle in the shot put and the optimum takeoff angle 
in the long jump.

The present study demonstrates 
that simplified analytical models can provide valuable physical insight 
into the optimum-angle problem in sports. 
Although the models considered here are intentionally simple, 
they successfully capture the essential mechanisms 
governing the optimum projection angle. 
The analytical framework developed in this study is expected to serve 
as a useful basis for understanding more realistic projectile motions 
involving aerodynamic forces and run-up.

\section*{Author's Note}

ChatGPT was used to assist with translation and English language refinement.



\begin{thebibliography}{99}
  
\bibitem[Clanet, 2015]{Clanet}Clanet, C., 
``Sports Ballistics'',
Annual Review of Fluid Mechanics, 47 (2015), 455–478, 
DOI:10.1146/annurev-fluid-010313-141255.

\bibitem[Cohen \textit{et al.}, 2013]{Cohen}Cohen, C., Darbois-Texier, B., Dupeux, G., Quéré, D. and Clanet, C., 
``The aerodynamic wall”, Proceedings of the Royal Society A, 470 (2013), 
20130497, DOI:10.1098/rspa.2013.0497.

\bibitem[de Mestre, 1991]{de Mestre}de Mestre, N., 
``The mathematics of projectiles in sport,''
Cambridge University Press, 1991.

\bibitem[Linthorne, 2001]{Linthorne_sp}Linthorne, N. P., 
``Optimum release angle in the shot put'', Journal of Sports Science, Vol. 19, (2001), pp. 359-372.

\bibitem[Linthorne, \textit{et al.}, 2005]{Linthorne_lj}Linthorne, N. P., Guzman, M. S., Bridgett, L. A., 
``Optimum take-off angle in the long jump'', Journal of Sports Science, Vol. 23, (2005), pp. 703-712.

\bibitem[Stewart, 2012]{Stewart}Stewart, S., M., 
``On the trajectories of projectiles depicted in early ballistic woodcut'', European Journal of Physics, Vol. 33, (2012), pp. 149-166.

\bibitem[Tait, 1890]{Tait}Tait, P. G., 
``Some points in the physics of golf'', Nature, Vol. 42, No. 1087 (1890), pp. 420-423.

\bibitem[Tsuboi, 2010]{Tsuboi_2010}Tsuboi, K., 
``A mathematical solution of the optimum takeoff angle in long jump'', 
Procedia Engineering, Vol. 2, Issue 2 (2010), pp. 3205-3210.

\bibitem[Tsuboi, 2012]{Tsuboi_2012} Tsuboi, K., 
``The Maximum Projection Angle of a Projectile under Drag and Lift'',
Transactions of the Japan Society of Mechanical Engineers, Part C, 
Vol. 78, No. 790 (2012), pp. 1972-1983 (in Japanese).

\bibitem[Tsuboi, 2015]{Tsuboi_2015} Tsuboi, K., 
``Optimum angle in projectile motion : Optimization problem in elementary mechanics'',
The Proceedings of Mechanical Engineering Congress Japan 2015, J2010203 (2015) 
(in Japanese).

\bibitem[Tsuboi, \textit{et al.}, 2015]{Tsuboi_jsiam_2015} 
Tsuboi, K., Hoshino, T. and Hamamatsu, Y., 
``Optimum Angle in Projection with Initial Speed Depending on Angle'', 
Transactions of the Japan Society for Industrial and Applied Mathematics, 
Vol. 25, No. 4 (2015), pp. 255–266 (in Japanese).

\bibitem[Tsuboi, 2025]{Tsuboi_2025} Tsuboi, K., 
``Projection Angles of Projectiles in Sports 
(Qualitative Assessment of the Effects of Aerodynamic Forces or Run-Up)'',
The Proceedings of Sports Engineering and Human Dynamics 2025, C000015 (2025) 
(in Japanese).

\end{thebibliography}
\end{document}